\documentclass[letterpaper,conference]{IEEEtran}
\IEEEoverridecommandlockouts

\usepackage{amsmath,amssymb,amsfonts}
\usepackage{algorithmic}
\usepackage{graphicx}
\usepackage{textcomp}
\usepackage{xcolor}
\usepackage{hyperref}
\hypersetup{
    colorlinks=true,
    linkcolor=blue,
    filecolor=magenta,      
    urlcolor=cyan,
}

\usepackage[
    style=ieee,
    doi=false,
    isbn=false,
    url=false,
    eprint=false,
    backend=bibtex,
    natbib=true
    ]{biblatex}
\bibliography{ref}

\def\BibTeX{{\rm B\kern-.05em{\sc i\kern-.025em b}\kern-.08em
    T\kern-.1667em\lower.7ex\hbox{E}\kern-.125emX}}
    
\begin{document}

\title{Humans prefer interacting with slow, less realistic butterfly simulations\\}

\author{Paige L. Reiter $^{1}$ \& Talia Y. Moore$^{1,2}$
\thanks{$^{1}$ Mechanical Engineering, Robotics Institute, University of Michigan, Ann Arbor, MI.
        {\tt\small plreiter@umich.edu}}
\thanks{$^{2}$ Robotics, Ecology and Evolutionary Biology, Museum of Zoology, University of Michigan, Ann Arbor, MI.
        {\tt\small taliaym@umich.edu}, ORCID: 0000-0003-0867-4512}
}

\maketitle
\pagestyle{empty}             
\thispagestyle{empty}

\begin{abstract}
How should zoomorphic, or bio-inspired, robots indicate to humans that interactions will be safe and fun?
Here, a survey is used to measure how human willingness to interact with a simulated butterfly robot is affected by different flight patterns.
Flapping frequency, flap to glide ratio, and flapping pattern were independently varied based on a literature review of butterfly and moth flight.
Human willingness to interact with these simulations and demographic information were self-reported via an online survey.
Low flapping frequency and greater proportion of gliding were preferred, and prior experience with butterflies strongly predicted greater interaction willingness.
The preferred flight parameters correspond to migrating butterfly flight patterns that are rarely directly observed by humans and do not correspond to the species that inspired the wing shape of the robot model.
The most realistic butterfly simulations were among the least preferred.
An analysis of animated butterflies in popular media revealed a convergence on slower, less realistic flight parameters.
This iterative and interactive artistic process provides a model for determining human preferences and identifying functional requirements of robots for human interaction.
Thus, the robotic design process can be streamlined by leveraging animated models and surveys prior to construction.
\end{abstract}


\section{Introduction}
\label{sec:intro}

Human interaction with animals and green spaces can have psychological and emotional benefits \cite{Cameron-Faulkner2018}.
For example, human perception of butterflies in outdoor environments is strongly associated with restorative and relaxing experiences \cite{Fuller2007,Pedretti2006,Soga2016}.
Because the frequency and outcomes of these situations are subject to chance, humans have developed lures --- 
birdfeeders, salt licks, nest boxes, etc. --- to increase the likelihood of a positive interaction.
However, lures are less effective in urbanized areas, where wild plants and animals are more rare \cite{B.Blair1997}.

\begin{figure}[t]
    \centering
    \includegraphics[width = \columnwidth]{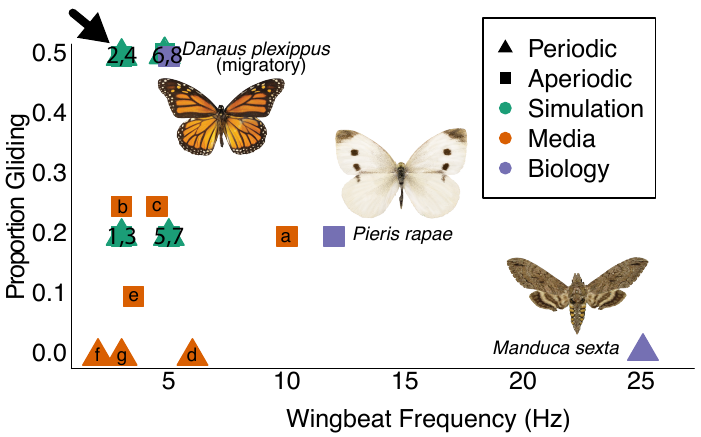}
    \caption{Comparison of flight characteristics for real butterflies (purple and images), animations of butterflies in popular media (orange), and butterfly robots simulated for this study (green). 
    The letters indicate the popular media source, corresponding to rows in Tab. \ref{tab:popref}.
    Note that in a) the character analyzed is a moth; all other examples are butterflies.
    The numbers indicate the video simulation, corresponding to trajectories in Fig. \ref{fig:unityscreenshot}.
    The preferred video of survey participants is indicated with a black arrow.
    Note that for simulated dataset, each specified combination of wingbeat frequency and proportion of gliding includes one aperiodic and one periodic datapoint.
    Images of the biological butterflies obtained from Yale Peabody Museum (\emph{Danaus plexippus}), Mus\'{e}um d'Histoire Naturelle de Toulouse (\emph{Pieris rapae}), and London (\emph{Manduca sexta}).
    The shape of each plotted data point indicates periodicity of the wingbeats.
    }
    \label{fig:shapes}
    \vspace{-.3cm}
\end{figure}

Zoomorphic, or bio-inspired, robots can be used to increase the likelihood of an interaction that is known to have positive emotional associations, such as a butterfly landing on a hand.
However, the positive emotional and psychological benefits of interacting with a butterfly-mimicking robot would be nullified if the combination of the motion and the appearance evoke an uncanny or repulsive effect.
Humans are repulsed by some robotic depictions of living organisms, an effect which is often referred to as the `Uncanny Valley' \cite{Mori1970}.
Such effects can arise from appearance and motion having different levels of abstraction, stylization, or anthropomorphism from their biological inspiration.
For humanoid animations, naturalistic movement tends to enhance acceptability of characters that are unacceptable when static, but distorted movement decreases the acceptability of realistic characters  \cite{Schwind2018}.
Animal-mimicking, or zoomorphic robots and animations can also evoke an uncanny effect \cite{Lo2020}, especially when anthropomorphism is combined with more realistic depictions \cite{Schwind2018}.
The extent to which previous research on the Uncanny Valley can be used to predict responses to invertebrate-mimicking robots is unknown because
invertebrate animals have less natural resemblance to human appearances and are less familiar to humans.

Butterflies are frequently depicted in television shows, movies, and video games to enhance the naturalistic ambiance of an environment \cite{Leskosky1988,Dika2008,Uhrig2018,Soga2016,Hogue1987}.
Because the success of artistic depictions depends on human emotional responses, the depictions of butterflies in popular media are likely to indicate human preference.
However, the extent to which popular animations are informed by biological data varies greatly \cite{Blumberg2003}, and there are no guidelines for assessing the likeability of zoomorphic animations based on the biological accuracy of their movement.

\begin{figure}[t]
    \centering
    \includegraphics[width = \columnwidth]{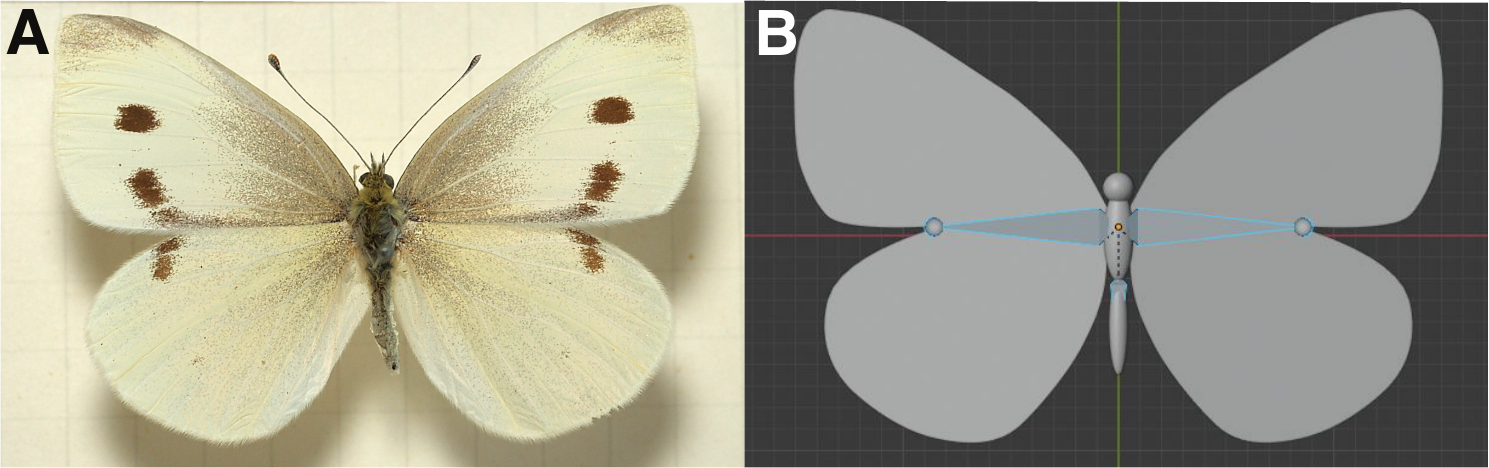}
    \caption{
    A) Female \emph{Pieris rapae} specimen (Photo by Sarefo (Wikimedia Commons) licensed under CC BY-SA 3.0).
B) Simple butterfly model in the Blender environment, with the armature used for animations outlined in blue.}
    \label{fig:butterfly}
    \vspace{-.3cm}
\end{figure}

With limited information regarding human preferences with respect to invertebrate motion, a simulated and interactive approach can direct the design of robots for positive interactions prior to fabrication.
Just as simulations can aid in optimizing terrestrial robotic locomotion for stability \cite{Paul2001}, human preference can be substituted as the goal  \cite{Bern2017}.

In this paper, aspects of real butterfly flight behaviors are reviewed (Sec. \ref{sec:template}) to inform flight simulations of a butterfly-mimicking robot (Sec. \ref{sec:model}).
The simulations are then used in a survey (Sec. \ref{sec:experiment}) to determine the effect of flight parameters on human willingness to interact with such a butterfly-mimicking robot.
The results from the survey (Sec.  \ref{sec:results}) are compared to depictions of butterflies in popular media (Sec. \ref{sec:popculture}).
The results of this survey demonstrate that precise biomimicry is not always the most effective strategy for designing zoomorphic robots, and further conclusions are drawn in Section \ref{sec:conclusion}.
Instead, by analyzing user feedback to simulations, human preferences for specific motions, and the corresponding functional capabilities of a robot, can be identified prior to fabrication.

\section{Biological Template}
\label{sec:template}
Butterflies and moths are flying insects of the order Lepidoptera, which contains over $175,000$ species. 
Within Lepidoptera, wing shape and size, body size, and flight characteristics all vary greatly.
In fact, if we consider a theoretical morphospace in which wing shape, wing size with respect to body, and body size represent independent axes, lepidopteran insects have evolved representatives across the entire available morphospace.
Because wing shape determines biomechanical capabilities, this morphological diversity implies that there are many different types of flight characteristics that would be considered realistic.
Thus, the design of butterfly-inspired robots should likely be informed by the intended function (in this case, human preference) and the limitations of the robotic system, rather than a narrow range of realistic movements.

Butterflies are some of the most popular insects in the world, likely due to their bright coloration, role in pollination, and lack of stinging or biting features \cite{Kellert1993, Lemelin2016}.
On the other hand, moths are often considered undesirable pests \cite{Kellert1993}.
Non-experts are easily able to distinguish between these groups and quickly decide whether to attempt an interaction \cite{Schlegel2015}.

Butterflies and moths may be distinguished by examining morphology and flight characteristics (Fig. \ref{fig:shapes}).
Butterflies generally have large forewings and hindwings with respect to their small body size.
For example, the large forewings of monarch butterflies (\emph{Danaus plexippus}) enable their long-range, low-energy $3000$ mile gliding migrations between Canada and Mexico.
During their cross-continental journey, monarchs spend over $80\%$ of their time gliding \cite{Ancel2017a}. 
Most humans generally do not observe the high-altitude migratory flight of monarch butterflies. 
Instead, we are much more likely to encounter butterflies like the cabbage white (\emph{Pieris rapae}), which exists in over 32 countries on every continent except South America and Antarctica \cite{Ryan2019}.
Cabbage white butterflies have average wingbeat frequencies between $10$ Hz and $15$ Hz and spend about $20\%$ of their time gliding \cite{Ha2013}.
On the other end of the spectrum, hawk-moth (\emph{Manduca sexta}) wings have a very high aspect ratio and are relatively small with respect to their large body size.
Hawk-moths are capable of incredibly fast, high-energy flapping flight, sometimes reaching speeds over $5.3$ m/s. 
Moths hardly ever glide, instead hovering in place at low speeds for precise tasks, such as drinking nectar from flowers. 
Furthermore, migratory monarchs can flap with wingbeat frequencies as low as $3$ Hz to conserve energy, while many moths flap at frequencies over $24$ Hz as they travel with rapid horizontal speeds \cite{Ancel2017a}, \cite{Willmott1997}.
These flapping frequencies are within the range of frequencies that can be distinguished by the human eye \cite{Heck1957,Harrington1981}.
Data regarding the wingbeat frequency, approximate flap to glide ratio, and wingbeat pattern were gathered from previous biomechanical analyses and video examination.
Because wing shape may be difficult to distinguish during flapping flight, the proportion of gliding and flapping and the wingbeat frequency are simple metrics that are likely used to distinguish butterflies from moths. 

\section{Simulated Model\label{sec:model}}

\begin{figure}[t]
    \centering
    \includegraphics[width = \columnwidth]{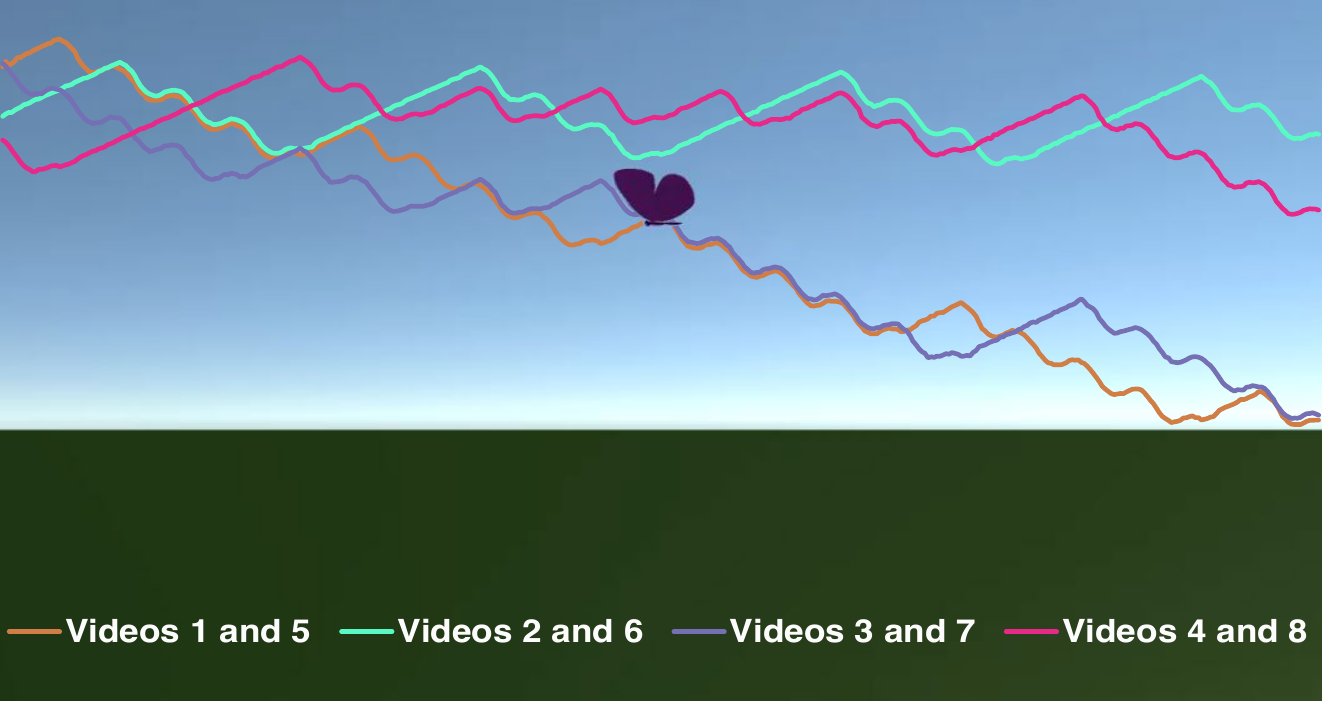}
    \caption{A screenshot from a butterfly simulation that was presented to participants, with the trajectories of all butterfly trials overlaid. }
    \label{fig:unityscreenshot}
    \vspace{-.5cm}
\end{figure}

\subsection{Appearance of butterfly model\label{sec:blend}}

The basic shape of a butterfly was modeled in Blender \cite{Blender} based on
the female \emph{Pieris rapae} (Fig. \ref{fig:butterfly}A), which was selected due to its ubiquity across several continents \cite{Ryan2019}.
Splines were used to trace the basic outline of the wings, and these splines were extruded to make thin 3D meshes.
Basic spheres were stretched into the shapes of the head, abdomen, and thorax to match the proportions of the butterfly template.
Because anthropomorphism has been demonstrated to produce diverging responses within a participant group \cite{Geerdts2014}, human-like stylization was not explored in this study.

These shapes were attached to a simple armature to generate the animations (Fig. \ref{fig:butterfly} B). 
Because the motion of the wings and the overall flight trajectory were the focus of this particular study,
the head, abdomen, and thorax were held fixed throughout the animated flights. 
Simple keyframes were used to create a looping 1 second animation of a 3Hz flapping motion, a 5Hz flapping motion, and a wings-extended gliding motion.
These simple animations and the model of simple shapes were exported from Blender and imported into Unity.

\subsection{Simulation of butterfly flight \label{sec:unity}}
The model and animation described in Section \ref{sec:blend} were imported into Unity \cite{Unity}.
C\# scripts were written to control the motion of the game objects in Unity.
The butterfly is rendered in purple and shown against a simple blue and green background representing the sky and ground, respectively (Fig. \ref{fig:unityscreenshot}).
The simple visuals were used to provide an abstracted version of a naturalistic background so as not to distract from the simulated butterfly.%

\subsection{Simulated parameters}

Based on the biomechanical features described
in Section \ref{sec:template}, the following parameters were independently varied in this study: wingbeat frequency,  proportion of gliding, and the pattern of wingbeats.
For each parameter, two values within the natural ranges of butterfly movement were selected to present during the videos for comparison (Tab. \ref{tab:parameters}). 
The low option for wingbeat frequency was set at $3$ Hz, and the high option was $5$ Hz, which is
the highest frequency renderable before the video outputs fail to generate smooth flapping motion.
The difference between the frequencies is noticeable by humans \cite{Heck1957,Harrington1981}.

Random motions are often added to repetitive movements to make animations appear more realistic and believable \cite{,Lee2002} and likely would have an effect on human willingness to interact with the robot.
Thus, periodic patterns were compared with aperiodic patterns of flapping and gliding.
Periodic simulations consisted of repeating patterns of either $4:1$ or $1:1$ ratios of flapping to gliding durations.
For aperiodic flapping, although the flapping to gliding ratios of the entire videos were either $4:1$ or $1:1$, the duration of each bout of flapping was determined by randomly obtaining $50$ samples, with replacement, from the set of numbers from $0$ to $9$, inclusive.
Because only a small number of samples were required to produce these simulations, resampling the numbers for each video would result in trajectories that likely differed greatly. 
Therefore, the same set of sampled numbers was used for all trials with aperiodic wingbeat patterns.
To alter flap to glide ratio of trials with aperiodic wingbeat patterns, the sampled numbers were assigned to different categories.
To produce a $4:1$ flap to glide ratio, the gliding motion was associated with values of $0$ to $1$ and the flapping motion was associated with values of $2$ to $9$.
For the $1:1$ flap to glide ratio, the gliding motion was associated with values of $0$ to $4$, while the flapping motion was associated with values of $5$ to $9$.
The resulting trajectories are shown in Figure \ref{fig:unityscreenshot}.

\begin{table}[t]
    \centering
    \begin{tabular}{c|c c c}
         Parameters &  Frequency & Pattern & Flap to Glide Ratio \\
         \hline
         Video 1 & $3$ Hz & Periodic & $4:1$ \\
         Video 2 & $3$ Hz & Periodic & $1:1$ \\
         Video 3 & $3$ Hz & Aperiodic & $4:1$ \\
         Video 4 & $3$ Hz & Aperiodic & $1:1$ \\
         Video 5 & $5$ Hz & Periodic & $4:1$ \\
         Video 6 & $5$ Hz & Periodic & $1:1$ \\
         Video 7 & $5$ Hz & Aperiodic & $4:1$ \\
         Video 8 & $5$ Hz & Aperiodic & $1:1$ \\
         \\
    \end{tabular}
    \caption{The flight parameters specified for each animated video.}
    \label{tab:parameters}
    \vspace{-.5cm}
\end{table}

While the aforementioned factors were varied, other parameters that might vary in nature were held constant in all videos.
The horizontal speed was held constant at $3.89$ body lengths per second.
To prevent distracting violations of physics, a vertical speed was specified for each wing flap ($1.55$ body lengths per second) and glide period ($- 1.30$ body lengths per second).
The butterfly moved within a plane perpendicular to the viewer, thereby maintaining a constant distance from the viewer.
The butterfly did not land at any point during any video, and it was consistently illuminated with diffuse lighting as it traveled across the scene.
No audio tracks were included in any videos.
Videos are posted in a playlist at \url{https://bit.ly/33ZSYn3}.

\section{Experimental Methods}
\label{sec:experiment}
A survey was created in Qualtrics \cite{Qualtrics} to determine whether biomechanical parameters influence the perceived friendliness of butterflies.
The University of Michigan Institutional Review Board determined that the study was exempt from review (HUM$00184754$).
The survey was primarily distributed via one email to the faculty, staff, and students associated with the University of Michigan's Robotics Institute and the UM Ecology and Evolutionary Biology department.
A few responses came from secondary contacts to members within those groups.
Though the survey was anonymous, participants had the option of listing their affiliation, age, familiarity with butterflies, and global regional locations where they had spent at least five years of their life.
The core portion of the survey involved participants watching eight video animations and responding to the following question for each:  ``Would you reach your hand out to let this butterfly land on you?''
Responses were reported on a scale from $0$ to $7$, in which an answer of $0$ corresponds to ``No thank you,'' while an answer of $7$ corresponds to ``I hope it lands on me!''
These values are reported hereafter as Interaction Willingness (IW) scores.
The order in which the videos were presented for each participant was randomized, but all videos were available at the same time, so participants could view the videos as many times and in whatever order they chose.
Lastly, participants had the option to leave a short answer suggestion for what they would change to make the butterfly appear friendlier.
There was no time limit to complete the survey.
The survey was open for seven days and received $131$ responses.
The survey is posted online at \url{https://bit.ly/2H2n68L}.

\section{Experimental Results and Discussion}
\label{sec:results}
\subsection{Human preferences}
Overall, the eight butterfly simulations were perceived as friendly, receiving an average IW rating of $5.07$ overall on a scale from $0$ to $7$. 
The overall favorite butterfly motion was featured in Video 2, which corresponded to $3$ Hz frequency wingbeat, a prescribed wingbeat pattern, and a $1:1$ ratio between time spent gliding and time spent flying (Tab. \ref{tab:results}).
In fact, the three most highly ranked videos share a wingbeat frequency of $3$ Hz.
The difference in IW between the highest ranked video (2) and the lowest ranked video (5) was highly significant (p = $1.67 e^{-4}$)

\begin{table}[t]
\vspace{1mm}
    \centering
    \begin{tabular}{c|c c c}
           &  Mean(IW) & SD(IW) & Median(IW) \\
         \hline
         Video 1 & $5.16$ & $1.85$ & $6$\\
         Video 2 & $5.36$ & $1.74$ & $6$\\
         Video 3 & $5.23$ & $1.77$ & $5$\\
         Video 4 & $5.05$ & $1.88$ & $5$\\
         Video 5 & $4.78$ & $1.88$ & $5$\\
         Video 6 & $5.05$ & $1.84$ & $5$\\
         Video 7 & $4.85$ & $1.90$ & $5$\\
         Video 8 & $5.06$ & $1.86$ & $5$\\
         \\
    \end{tabular}
    \caption{A summary of Interaction Willingness (IW) scores for each video}
    \label{tab:results}
    \vspace{-.5cm}
\end{table}

While all of the parameters shown in the videos were possible for real butterflies, participants demonstrated a strong preference for the very slow wingbeat frequency of $3$ Hz (p = $4.97 e^{-5}$).
Participants also demonstrated a weak preference for more gliding (p = $0.0730$). 
Unless a butterfly is actively descending, riding an air current like they might during a long migratory event, or taking advantage of ground effects, butterflies are very unlikely to have an even distribution of flapping and gliding time.
Unexpectedly, wingbeat pattern did not have a significant effect on IW scores (p = $0.489$).
Overall, the videos with the highest IW scores display flight parameters corresponding to migratory flights above the tree canopy, rather than flying through gardens at approximately human height.

These surprising results prompted further analysis to investigate whether demographic factors informed these preferences.

\subsection{Demographic Factors}

Although butterflies are globally distributed and generally considered positively by many cultures, there is a chance that participants may be more familiar with static representations of butterflies as decorative motifs if they have not had personal encounters.
There may also be differences among individuals who generally like or dislike all butterflies, versus those who are more discerning among butterflies.
Thus, participant responses to several optional questions regarding demographic information (e.g. age, experience with butterflies, academic affiliation) were analyzed to determine their effect on IW scores.
A participant's level of discerning among butterfly simulations was evaluated by measuring the range of their IW scores, while overall affinity was evaluated by measuring the mean.

\begin{figure}[t]
    \centering
    \includegraphics[width=\columnwidth]{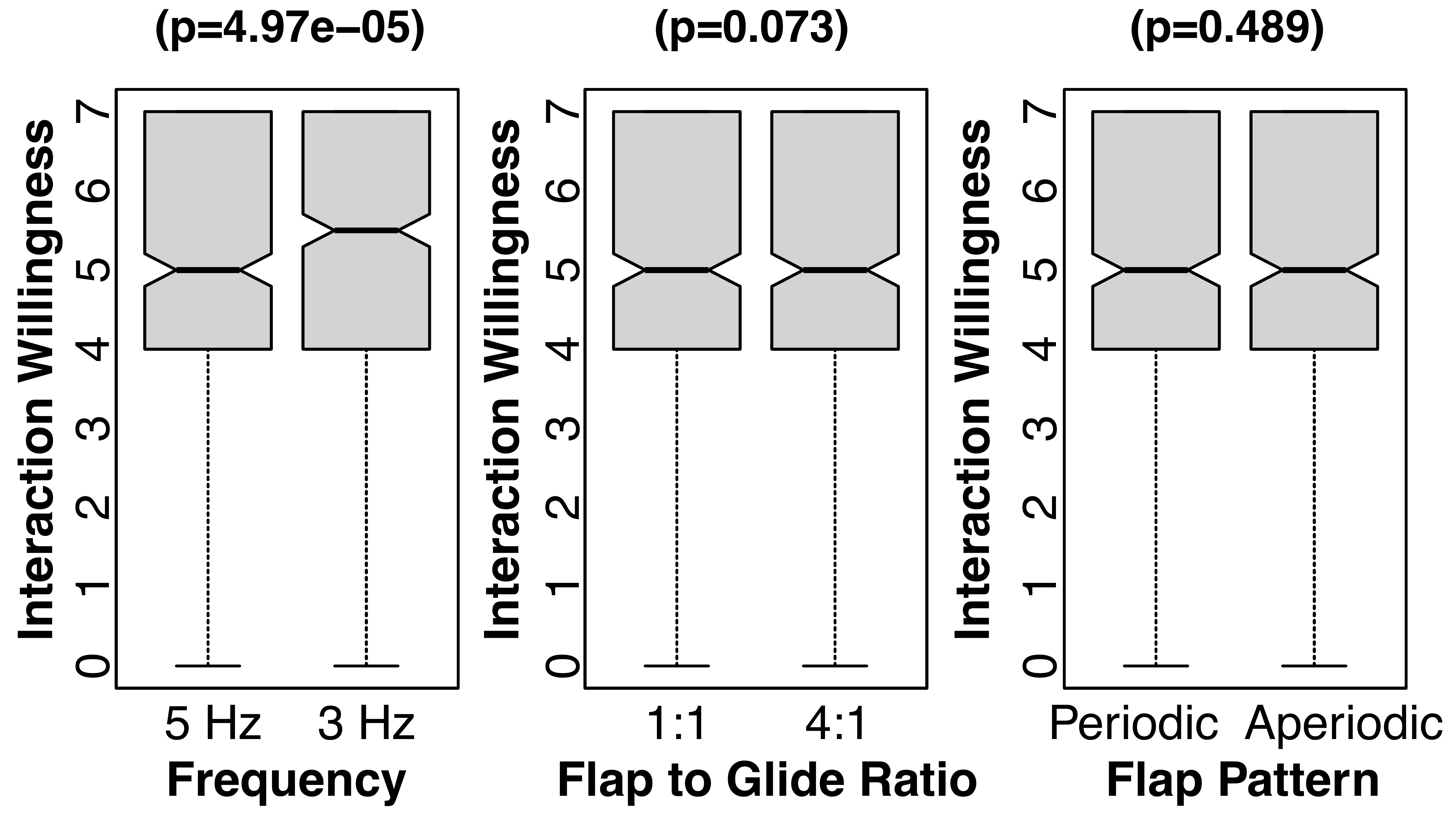}
    \caption{Survey participant willingness to interact varied with respect to different flight parameters.}
    \label{fig:biomech}
    \vspace{-.2cm}
\end{figure}

\begin{figure*}[t]
    \centering
    \includegraphics[width = \textwidth]{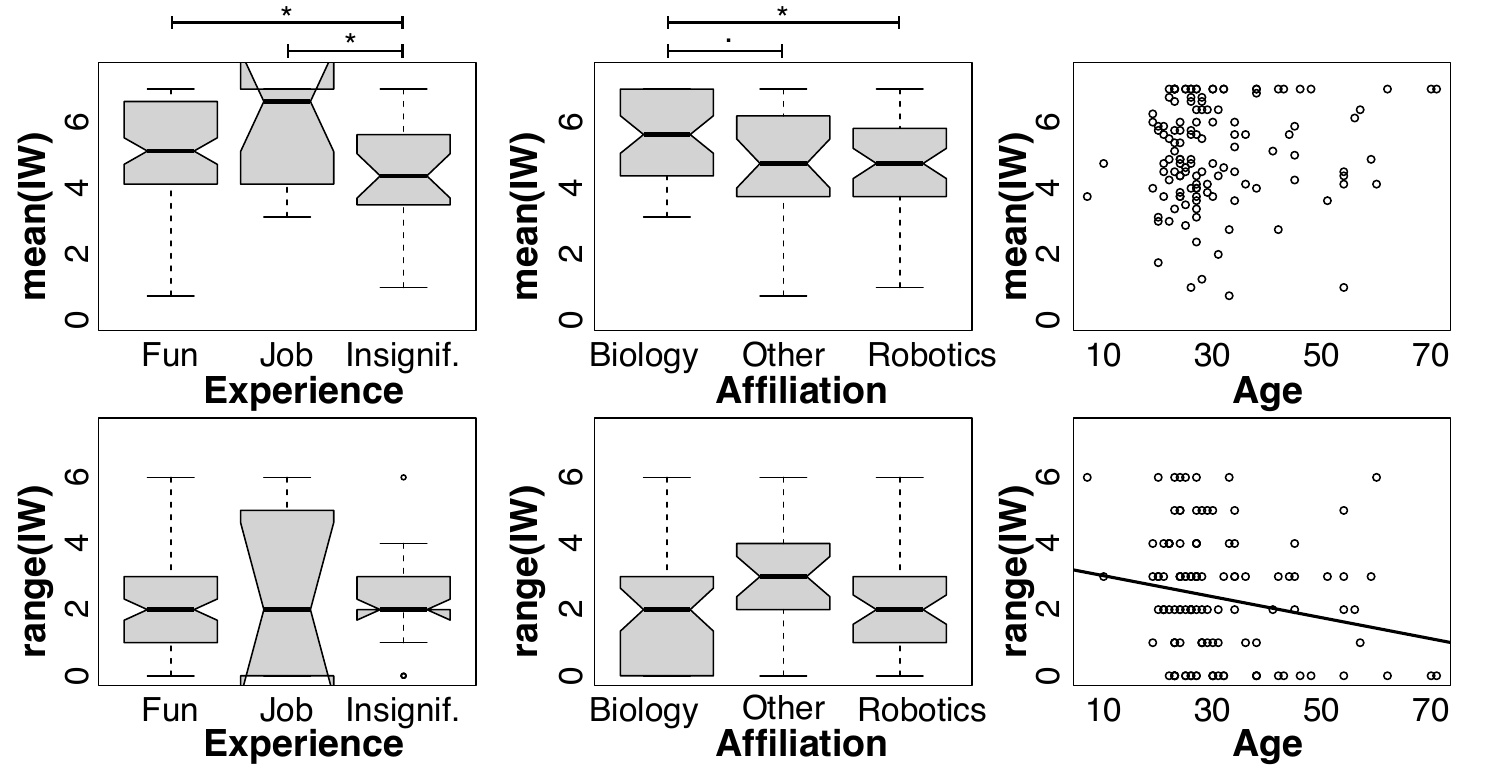}
    \caption{Comparison of Interaction Willingness (IW) with respect to demographic factors.
    Asterisks indicate significance level $p < 0.05$.
    A period indicates significance level $0.05 < p < 0.1$.
    Academic affiliation and previous experience with butterflies significantly predicted mean IW scores for each participant.
    The range of IW scores for a participant decreased with age, but this trend did not explain the majority of the variation in the data.
    The y-axes were truncated at the maximum available score of seven.}
    \label{fig:demographics}
\end{figure*}

Participants who interact with butterflies as part of their job (n = $9$) demonstrated highest overall affinity for butterfly videos and the widest range of discerning.
Participants who have not meaningfully interacted with butterflies (n = $25$) scored the videos significantly lower than those who have interacted with butterflies for fun (n = $97$, p = $0.0302$), and those who interact with butterflies as a part of their job (p = $0.0450$).
The mean level of discerning was not significantly determined by affiliation. 
Although this option was available, no participants reported zero familiarity with butterflies.

Participants affiliated with the Ecology and Evolutionary Biology department (n = $53$) showed the highest overall and most consistent IW scores regardless of previous experience with butterflies.
Biologists rated videos significantly higher than participants affiliated with the Robotics Institute (n = $51$, p = $5.22 e^{-3}$) and somewhat higher than those with Other affiliations (n = $27$, p = $0.0759$).
All participants who interact with butterflies as a part of their jobs were affiliated with the Biology department, but biologists who do not study butterflies were the least discerning participants.
Roboticists who had interacted with butterflies for fun ranked videos higher on average than those who have had insignificant interaction with butterflies (p = $0.0256$), but all roboticists had similar levels of discerning.
Participants with Other affiliations who have interacted with butterflies for fun were almost as discerning as those who study butterflies as their occupation, but differences in discerning according to experience were not significant within this group.

Participant ages ranged from 7 to 71\footnote{Children were not solicited for participation in the survey, but multiple survey entries per household were accepted.}, with $50$\% of participants within the range of 24 to 34.
Age of participant did not significantly affect the overall affinity for butterflies (p = $0.2$).
Discerning among butterfly videos significantly decreased with age (p = $0.0218$), but this trend did not explain the variation in most of the data (Adjusted $R^{2}$ = $0.0337$). 

There were no interactions among the biomechanical (fixed effects) and demographic (random effects) when tested as a linear mixed-effects model using the lmer function in the lmerTest package for R \cite{lmerTest}.

\subsection{Open responses}
During the survey period, multiple participants directly contacted the researchers with positive messages, such as ``... thinking about robot butterflies made my day a little brighter.''
An analysis of the verbatim responses at the end of the survey brought to light some interesting topics.
$66$\% of the $131$ participants chose to leave a note at the end of the survey in the box asking for suggestions on how to make the butterflies appear even more friendly.
$26$ participants explicitly suggested alterations to the static visual appearance of the butterfly, asking researchers to make the robot look like a familiar monarch species or to add anthropomorphic eyes.
The butterfly in this study was intentionally designed to be nondescript so that positive associations with a specific species due to previous interactions would be equally unlikely for all participants, regardless of regional origin.
The effects of stylization and anthropomorphization can be examined in future studies.

$54$ participants made suggestions related to behavior, which was broken down further into several categories.
$15$ participants explicitly described the flapping motion, including requests to increase or decrease the overall speed or adding variability in the wingbeat frequency.
$15$ participants described the gliding behavior, with some calling for more gliding and others calling for less.
A few participants questioned the physics of such a rapid descent associated with gliding.
Lastly, $10$ participants addressed  the horizontal speed of the butterfly, which was a fixed variable across all trials.
Most participants suggested slowing butterflies to speeds that correspond to values below the range of realistic speeds.

$30$ suggestions described the overall flight trajectory.
These suggestions ranged from choosing a more simple trajectory to suggestions of more erratic trajectories.
$16$ participants requested more interaction with the butterfly. 
Several suggestions included alterations in trajectories and in interaction, and were counted in both.
$16$ participants described wanting to know whether the butterfly wanted to interact with them, suggesting either that the flight path come near or circle the observer, or that the flapping frequency decrease with proximity to the observer.
Although such indications of intent are potentially informed by interactions with humans, dogs, or cats instead of butterflies, the underlying concept of the robot giving the appearance of informed consent is a fascinating avenue for future research.

\section{Depictions of butterflies in popular media}
\label{sec:popculture}

\subsection{Data collection}
\begin{table*}[t]
\vspace{1mm}
    \centering
    \begin{tabular}{ll llll}
         &Popular Media Name &  Type & Year & Studio & Director/Creator \\
         \hline
         a&\emph{A Bug's Life} & Feature-length film & $1998$ & Pixar Animation Studios & John Lasseter \\
         b&\emph{The Road to El Dorado} & Feature-length film & $2000$ & DreamWorks Animation & Eric Bergeron and Don Paul \\
         c&\emph{Big Hero 6} & Feature-length film & $2014$ & Walt Disney Animation Studios & Don Hall and Chris Williams \\
         d&\emph{Miraculous: Tales of Ladybug and Cat Noir} & Television show & $2015$ & Zagtoon and Method Animation & Thomas Astruc, et al. \\
         e&\emph{Lovecraft Country} & Television show & $2020$ & Monkeypaw Productions, et al. & Cheryl Dunye \\
         f&\emph{Second Life} & Video game & $2003$ & Linden Lab & Hana Kozlowski\\
         g&\emph{Legend of Zelda:  Breath of the Wild} & Video game & $2017$ & Nintedo & Hidemaro Fujibayashi \\
    \end{tabular}
    \caption{The popular media sources of the animated butterflies analyzed for this study. \label{tab:popref}}
    \vspace{-.5cm}
\end{table*}

The authors recorded animated butterflies encountered during their normal interaction with video games, animated television shows, live-action television shows, and animated feature-length films during a four-month period, from June 11 to October 11, 2020 (see Tab. \ref{tab:popref}).
Wingbeat frequencies and patterns were determined by counting the number of video frames per wing beat, and the proportion of gliding was measured over the entire duration of the recorded flight.

\subsection{Trends in animated butterfly motion}
Butterflies depicted in popular media had wingbeat frequencies between 3 and 6 Hz, much like the animations created for the survey (Fig. \ref{fig:shapes}).
This is on the extremely low end for real-life butterflies, similar to the frequencies used by migrating butterflies that are rarely observed by casual viewers.
The depictions of butterflies in feature-length films and live-action television shows featured periods of gliding between the flaps, though the examples from video games and animated television shows did not include any gliding.
In the characterized films, the butterflies spent about $20\%$ to $25\%$ of their time gliding.
For the other media types, the lack of gliding may be a consequence of simplifying the animation process by limiting butterfly flight to a fully periodic pattern so that the same animations can be reused in multiple episodes of a television show or in different locations in a video game.
Feature-length films and live-action television shows were also more likely to feature aperiodic flapping cycles.

\section{Conclusion}
\label{sec:conclusion}
Based on the overall positive rankings of simulations, the results of this study suggest that butterfly-mimicking robots can facilitate positive interactions with humans.
Although the results were generally positive, they were not uniform, indicating that humans do have preferences for slower forms of butterfly motions.
Therefore, butterflies are an appropriate inspiration for zoomorphic robots, and tuning the flight behaviors is likely to have an effect on user acceptance and interaction.
This preference may be explained by an aversion to moths, which display a higher wingbeat frequency and greater proportion of flapping to gliding.
Alternatively, a preference for slower and more predictable motion may be driving this pattern.
Such `sensory bias' for a specific feature has been revealed to drive the evolution of extreme features in animals undergoing intersexual selection \cite{Fuller2005}.
Repeating this experiment using more abstract images, such as floating spheres, can help determine whether moth aversion or sensory bias drives this preference, and how these preferences compare to predictions based on the Uncanny Valley.

Our results provide foundational information for examining human affinity for representations of invertebrate animals as a function of their similarity to a real organism.
Unlike previous examinations of static vertebrate animal representations that revealed Uncanny Valley patterns of likeability with respect to verisimilitude \cite{Schwind2018,Lo2020}, 
humans in this study preferred motions that are slower than observed in real butterflies. 
To determine whether the preference patterns found in this study can be extended to all invertebrates, a range that spans human likeability, from spiders and centipedes to butterflies and ladybugs, could be examined.

Anthropomorphism may also influence the Uncanny Valley effect with respect to vertebrate animal representations \cite{Schwind2018}.
Because invertebrate animals have less morphological resemblance to humans, the effect anthropomorphism has on human preference involves more suspension of disbelief and is therefore difficult to predict.
Furthermore, examining the response of children to picture books with or without anthropomorphism demonstrated diverging effects in boys and girls \cite{Geerdts2014}.
Thus, future studies should carefully consider whether and when anthropomorphism should be applied to invertebrate-mimicking robots.

This study also adds to a growing body of work examining the effect of movement on the Uncanny Valley.
In humanoid representations, realistic movement has been shown to rehabilitate statically unfamiliar representations, but distorted movement decreases the familiarity of statically realistic representations \cite{Castro-Gonzalez2016,Piwek2014}.
Here, although static and dynamic representations were not compared, variation in movement patterns associated with different proportions of gliding had only a weak effect on human willingness to interact with the simulations.
Humans are likely attuned to sensing changes in trajectory randomness, as evidenced by targeting accuracy decreasing for more erratically moving targets \cite{Richardson2018,Clemente2016}.
Animals may also experience strong selective pressures favoring trajectory randomness for predator evasion \cite{Mooree,Combes2012,SkowronVolponi2018}.
Although the trajectories simulated in this study varied as a result of altering the flapping and gliding patterns, the trajectory randomness was not directly varied. 
In future research, studying the effect of random movements by varying trajectory entropy can precisely characterize how humans respond to erratic motion of zoomorphic robots.

The findings from this study converge with the representations of butterflies in animated media.
Interestingly, a survey of popular media indirectly indicates that slower and more predictable butterflies are aesthetically pleasing, even if they share less resemblance to real animal movement (Fig. \ref{fig:shapes}).
This phenomenon could potentially be explained by the fact that artistic animations often go through rounds of feedback and revision to shape their final product.
Here, we demonstrate how a similar interactive and iterative process can be used to design robotic systems for user interface.
Leveraging simulated motion is a useful strategy to decrease the effort required for each iteration, thereby streamlining the manufacturing process by determining functional requirements prior to robot construction \cite{Bern2017}.

The feedback from the participants, especially the open responses, suggest how future interactive and simulated studies can provide additional guidance for the design and deployment of such a robotic system.
An important takeaway from this study is that demographic factors strongly determined human willingness to interact with a butterfly robot.
While the generic visual appearance of the butterfly in this study was designed to ensure consistent levels of familiarity across cultural backgrounds, examination of regional preferences or interaction with the local setting can inform the design of distinct models for different target audiences.
As animals are also frequently represented in animation and robotics, examining the effect of their diverse movement patterns on human acceptance will likely benefit many different research fields and industries.

\section{Acknowledgements}
The authors thank E.L. Moore for providing footage and description of the animated butterfly in \emph{Second Life},
A. Reiter for assistance with simulations,
members of the EMBiR lab for providing feedback on a pilot version of the survey,
and members of the ROAHM Lab and J. Socha for providing feedback on the manuscript.




\renewcommand{\bibfont}{\normalfont\small}
{\renewcommand{\markboth}[2]{}
\printbibliography}

\end{document}